\documentclass[conference]{IEEEtran}
\IEEEoverridecommandlockouts

\usepackage{cite}
\usepackage{amsmath,amssymb,amsfonts}
\usepackage{graphicx}
\usepackage{textcomp}
\usepackage{xcolor}
\usepackage{booktabs}
\usepackage{multirow}
\usepackage{adjustbox}
\usepackage{url}
\usepackage[most]{tcolorbox}
\usepackage{hyperref}
\usepackage{dblfloatfix}
\usepackage{svg}
\usepackage{graphicx,amsmath}
\begin{document}

\title{Unified Shared Encoder in Spoof-Aware Speaker Verification with Hybrid Wavelet Prompt Tuning}

\author{\IEEEauthorblockN{Aref Farhadipour$^{1}$, Srikanth Madikeri$^{1}$, Teodora Vukovic$^{1}$, Volker Dellwo$^{1}$, Petr Motlicek$^{2,3}$}
\IEEEauthorblockA{$^{1}$Department of Computational Linguistics, University of Zurich, Switzerland \\ $^{2}$Idiap Research Institute, Switzerland \qquad
$^{3}$Brno University of Technology, Czech Republic\\
\texttt{aref.farhadipour@uzh.ch}\\
}
}

\maketitle

\begin{abstract}
Spoofing-aware speaker verification (SASV) must confirm who is speaking and that the speech is genuine, but current systems gain one at the expense of the other. Modular cascades are accurate yet run two large self-supervised encoders, whereas end-to-end models are efficient but lose accuracy when a single embedding has to serve two conflicting objectives. We show that this trade-off between efficiency and specialization is avoidable. A single frozen W2V-BERT~2.0 backbone is adapted per task by deep hybrid Wavelet Prompt Tuning (WPT), so each branch obtains its own view of the shared encoder through dedicated prompts and a task head, and the scores are combined only at inference. Training only about 7M parameters, a small fraction of those a strong two-encoder cascade requires, the system surpasses it on SpoofCeleb evaluation with 0.03\% CM-EER and 0.038 min a-DCF against 0.16\% and 0.047. Because the backbone is shared and frozen, new branches attach without retraining the heads and prompts already in place. On ASVspoof5, with adversarial attacks and codec distortions, it reaches 0.092 min a-DCF and 3.60\% CM-EER using only the provided data, indicating that the design holds under realistic in-the-wild threats.
\end{abstract}

%

\section{Introduction}
\label{sec:intro}

The rapid rise of generative artificial intelligence has fundamentally altered the landscape of biometric security. While Automatic Speaker Verification (ASV) systems have reached high levels of precision in distinguishing between target and non-target speakers \cite{far2024analysis}, they remain critically vulnerable to spoofing attacks, including high-fidelity text-to-speech (TTS) \cite{tan2021survey}, voice conversion (VC) \cite{gupta2025direct}, and sophisticated neural deepfakes \cite{li2025survey}. To mitigate these threats, Spoofing-Aware Speaker Verification (SASV) has emerged as a critical paradigm, designed to unify ASV and countermeasure (CM) subsystems into an integrated robust defense.

Traditional SASV approaches have primarily relied on modular cascade architectures composed of independent ASV and CM subsystems \cite{peng2025but}, where each utterance is first processed by the CM module and, based on its decision, is either passed to the ASV module or directly assigned a final rejection score. While these cascaded approaches preserve task modularity, they suffer from significant computational redundancy by maintaining two separate large-scale backbones to process the same input \cite{far2026spoofing}. Such two-backbone designs are also hard to extend, since adding a task or data source requires a further backbone rather than a lightweight module. As noted by Kurnaz et al. \cite{kurnaz2025joint}, fusing the scores of two \emph{independently optimized} subsystems, whose internal representations never interact, frequently fails to capture the complex dependencies between identity and liveness. This lack of representation-level interaction often results in detrimental error propagation, where inaccuracies in one module negatively impact the global system decision.

Independently of how the two tasks are ultimately combined, the quality of the underlying representation is decisive for both. Consequently, the extraction of high-dimensional acoustic representations via large-scale pre-trained Self-Supervised Learning (SSL) models, most notably WavLM \cite{chen2022wavlm} and Wav2Vec 2.0 \cite{baevski2020wav2vec}, has become a standard approach in the field. For example, \cite{peng2025but} integrates two different SSL models with a Multi-Head Factorized Attention (MHFA) backend \cite{peng2023attention} for ASV and an AASIST back-end \cite{jung2022aasist} for CM. While effective, these systems typically require complex multi-stage training or several SSL backbones, leading to prohibitive computational costs \cite{far2026spoofing}.

To overcome these limitations, recent research has pivoted toward End-to-End (E2E) architectures, where identity and liveness cues are extracted simultaneously within a single model to foster cross-task synergy \cite{wang2024asvspoof}.
However, a primary shortcoming of this strategy is that these models often struggle to achieve competitive performance on the SASV task compared to modular baselines. This performance gap is frequently attributed to the challenge of optimizing a single representation for two conflicting objectives.


To resolve this efficiency--specialization trade-off, we propose an SASV system that employs the Unified Shared-Encoder (USE) architecture with specialized task-specific heads. Specifically, our framework utilizes a W2V-BERT 2.0 backbone \cite{barrault2023seamless} to extract a diversity of hierarchical features from the input audio, which are then processed by parallel ASV and CM heads conditioned by parameter-efficient hybrid Wavelet Prompt Tuning (WPT) \cite{lester2021power} while keeping the backbone strictly frozen. WaveSP-Net \cite{xuan2025wavesp} recently fused prompt tuning with wavelet-domain transforms for speech deepfake detection on a frozen XLSR, an idea we adopt and extend to a shared encoder serving both ASV and CM. We select W2V-BERT 2.0 because its Conformer blocks capture both local acoustic detail and long-range context, and its multilingual pre-training on 4.5M hours provides the broad, general-purpose features that a single shared encoder needs to serve two tasks at once. We adapt it via prompt tuning rather than LoRA, adapters, or full fine-tuning because prompts leave every backbone weight untouched, which is exactly what allows multiple task-specific branches to share one frozen encoder and be attached independently. Crucially, unlike E2E systems that force a \emph{single} shared embedding to satisfy both objectives, USE shares only the frozen, general-purpose knowledge of the encoder while assigning each task its own prompt set and head. The task-specific prompts provide distinct, conditioned virtual context over the same backbone for ASV and CM, so as to sidestep the single-representation conflict without incurring the cost of two separate backbones. Our system is therefore not an end-to-end single-embedding model but a shared-encoder multi-task framework whose per-task scores are combined at the score level. 
This shared encoder strategy facilitates seamless architectural scalability, allowing for the independent integration of new task-specific heads without necessitating the retraining or modification of existing modules.

As spoofing algorithms become more advanced \cite{xie2025detect}, they increasingly mimic the spectral and prosodic characteristics of the target speaker, rendering single-layer representations insufficient \cite{peng2025but}. Our approach therefore aggregates features from all transformer layers, giving both the ASV and CM heads access to a comprehensive hierarchy that spans fine-grained acoustic artifacts and high-level identity cues. We validate this design on two challenging in-the-wild benchmarks, SpoofCeleb \cite{jung2025spoofceleb} and ASVspoof5 \cite{wang2024asvspoof}, which stress robustness through environmental noise, reverberation, and diverse codec and channel distortions.

\begin{figure}[t]
    \centering
    \includegraphics[angle=90, scale=1]{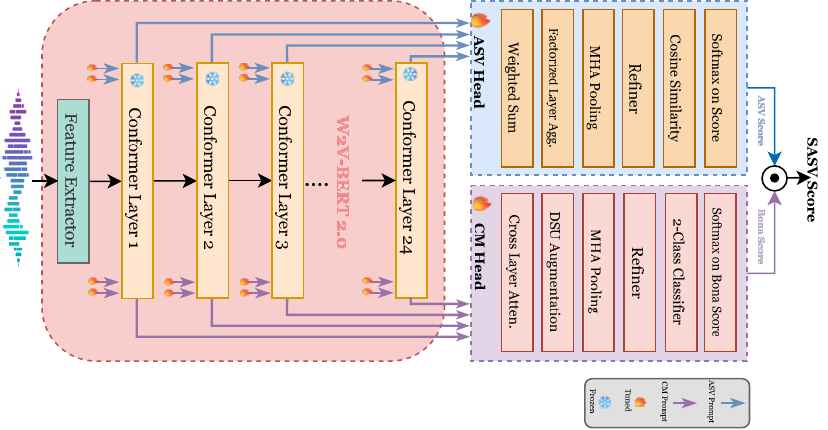}
    \caption{Proposed USE architecture for the SASV task. At each transformer layer, two arrows of the same color denote the simultaneous injection of regular and wavelet prompt sets.}
    \label{fig:sasv_diag}
\end{figure}

\section{Proposed Method}
\label{sec:method}

The proposed SASV framework, illustrated in Figure \ref{fig:sasv_diag}, adopts an integrated shared-encoder architecture designed to unify speaker verification and spoof detection within a single processing pipeline. The core of our system is the W2V-BERT 2.0 backbone, which comprises 24 Conformer layers and approximately 600M parameters, having been pre-trained on 4.5 million hours of diverse speech data \cite{barrault2023seamless}.
W2V-BERT 2.0 serves as a foundation for this task due to its self-supervised representations learned from multilingual pre-training. Its hybrid training strategy combines complementary learning objectives within a Conformer-based architecture, providing multi-layer features that support feature aggregation.

The proposed framework utilizes parameter-efficient prompt tuning to adapt the frozen W2V-BERT 2.0 backbone. Unlike full-model adaptation, prompt tuning introduces a set of learnable tokens. 
In the context of our 24-layer Conformer-based backbone, these tokens are defined as a learnable matrix $\mathbf{P} \in \mathbb{R}^{L \times D}$, where $L$ denotes the total number of prompts and ($D=1024$) matches the model's hidden dimension. These tokens serve as learnable task-specific priors that are prepended to the input sequence of each Transformer layer. During the self-attention operation, these prompts are projected into the Key and Value matrices, providing a virtual context for the audio-derived Queries. This mechanism is designed to guide the frozen backbone's attention toward task-relevant features without modifying the underlying pre-trained weights (Section~\ref{subsec:ablation}).

\subsection{Hybrid Wavelet Prompt Tuning}
\label{subsec:wpt}

Let the frozen backbone be a stack of $N=24$ Conformer layers $\{f_\ell\}_{\ell=1}^{N}$ operating on a hidden sequence of dimension $D=1024$, and let $\mathbf{H}_{\ell-1}\in\mathbb{R}^{T\times D}$ denote the audio hidden states entering layer $\ell$, where $T$ is the number of frames. For each layer we attach two task-specific, learnable prompt banks: a \emph{regular} prompt $\mathbf{P}^{r}_{\ell}\in\mathbb{R}^{L_r\times D}$ and a \emph{wavelet} prompt $\mathbf{P}^{w}_{\ell}\in\mathbb{R}^{L_w\times D}$, with $L_r=15$ and $L_w=8$ tokens, selected by hyperparameter tuning on dev data. Both banks are randomly initialized and constitute the only trainable quantities of the prompt module, while all backbone weights remain frozen.

\noindent\textbf{Wavelet transformation.} Rather than injecting the Wavelet bank directly, at every forward pass we first pass it through a single-level ($J{=}1$) two-dimensional Haar Discrete Wavelet Transform (DWT) $\mathcal{W}(\cdot)$. Treating $\mathbf{P}^{w}_{\ell}$ as a $2$D token--channel map, the transform decomposes it into one low-pass approximation sub-band $\mathbf{A}_\ell$ and three high-pass detail sub-bands $\{\mathbf{D}^{\text{LH}}_\ell, \mathbf{D}^{\text{HL}}_\ell, \mathbf{D}^{\text{HH}}_\ell\}$:
\begin{equation}
\big(\mathbf{A}_\ell,\ \mathbf{D}^{\text{LH}}_\ell,\ \mathbf{D}^{\text{HL}}_\ell,\ \mathbf{D}^{\text{HH}}_\ell\big) = \mathrm{Haar}\big(\mathbf{P}^{w}_{\ell}\big),
\end{equation}
which are concatenated along the sub-band axis and reshaped back to the original prompt size, yielding
\begin{equation}
\tilde{\mathbf{P}}^{w}_{\ell} = \mathcal{W}\big(\mathbf{P}^{w}_{\ell}\big) \in \mathbb{R}^{L_w \times D},
\end{equation}
where $\mathcal{W}(\cdot)$ denotes the Haar DWT followed by sub-band concatenation and reshaping.
The transform operates on the prompt-token $\times$ channel matrix $\mathbf{P}^{w}_{\ell}$ (not on audio time/frequency), injecting a multi-resolution prior into prompt tokens seen only as frozen Key/Value context. Standard soft prompts mainly learn global task-specific shifts \cite{lester2021power,xuan2025wavesp}. Because synthesis artifacts often appear as localized high-frequency perturbations \cite{gao2021generalized,xie2025detect}, the wavelet basis provides an inductive multi-resolution prior to guide attention, while regular prompts maintain flexible adaptation.

\noindent\textbf{Deep prompt injection.} The two banks are prepended to the audio states at the input of \emph{every} layer (deep prompting). Denoting token-axis concatenation by $[\,\cdot\,;\,\cdot\,]$, $P=L_w+L_r$, and $M=P+T$, the augmented input to layer $\ell$ is
\begin{equation}
\hat{\mathbf{H}}_{\ell-1} = [\tilde{\mathbf{P}}^{w}_{\ell}; \mathbf{P}^{r}_{\ell}; \mathbf{H}_{\ell-1}] \in \mathbb{R}^{M \times D}.
\end{equation}
Within the frozen multi-head self-attention of $f_\ell$, only the audio frames query the prompt-augmented sequence. Using the frozen projections $\mathbf{W}_Q,\mathbf{W}_K,\mathbf{W}_V$ and per-head dimension $d_k$, we define
\begin{equation}
\mathbf{Q} = \mathbf{H}_{\ell-1}\mathbf{W}_Q, \quad \mathbf{K} = \hat{\mathbf{H}}_{\ell-1}\mathbf{W}_K, \quad \mathbf{V} = \hat{\mathbf{H}}_{\ell-1}\mathbf{W}_V,
\end{equation}
\begin{equation}
\mathrm{Attn}_\ell = \mathrm{softmax}\!\left(\frac{\mathbf{Q}\mathbf{K}^{\top}}{\sqrt{d_k}}\right)\mathbf{V},
\end{equation}
where $\mathbf{Q} \in \mathbb{R}^{T \times d_k}$ is formed from the audio states only, whereas $\mathbf{K},\mathbf{V} \in \mathbb{R}^{M \times d_k}$ include the prompt tokens.
After the Conformer block update, the first $P$ prompt positions are discarded and only the $T$ audio frames are propagated to the next layer:
\begin{equation}
\mathbf{H}_{\ell} = \big(f_\ell(\hat{\mathbf{H}}_{\ell-1})\big)_{P:M} \in \mathbb{R}^{T \times D}.
\end{equation}
Fresh layer-specific prompts are then re-injected at layer $\ell{+}1$.
Since only $\{\mathbf{P}^{r}_{\ell},\mathbf{P}^{w}_{\ell}\}_{\ell=1}^{N}$ are updated, the prompt module introduces just $N\,(L_r+L_w)\,D$ trainable parameters while guiding the attention of the frozen encoder toward task-discriminative cues.
Each task branch (prompts and head) is trained independently on its respective objective while keeping the 600M backbone frozen, and both are mounted onto the shared backbone at inference. Both heads read all 24 layers, apply attentive temporal pooling, pass through a Refiner bottleneck module, and map to a 256-dimensional embedding.

\subsection{ASV Head: Layer-Fused MHFA }

The ASV head extracts identity-discriminative embeddings from the full hierarchical depth of the frozen backbone using Multi-Head Factorized Attention (MHFA) \cite{peng2023attention}. We aggregate the 24 Transformer layers using a factorized weighted-sum strategy, where separate learnable weights are assigned to the key and value streams. This approach decouples attention-based conditioning from content aggregation. Denoting the $N{=}24$ frozen layer outputs by $\{\mathbf{H}^{(\ell)}\}_{\ell=1}^{N}$ with $\mathbf{H}^{(\ell)}\in\mathbb{R}^{T\times D}$, the fused key and value streams are
\begin{equation}
\bar{\mathbf{K}}=\sum_{\ell=1}^{N}\alpha^{k}_{\ell}\,\mathbf{H}^{(\ell)},\qquad
\bar{\mathbf{V}}=\sum_{\ell=1}^{N}\alpha^{v}_{\ell}\,\mathbf{H}^{(\ell)},
\end{equation}
where $\boldsymbol{\alpha}^{k}{=}\mathrm{softmax}(\mathbf{w}^{k})$ and $\boldsymbol{\alpha}^{v}{=}\mathrm{softmax}(\mathbf{w}^{v})$ are independent learnable layer weights for the key and value paths. We deliberately keep the ASV aggregation as a lightweight static weighted sum because speaker identity is a globally stable attribute. A fixed per-layer importance profile is therefore sufficient and yields a more stable, lower-variance embedding than input-dependent fusion.

The fused streams $\bar{\mathbf{K}},\bar{\mathbf{V}}\in\mathbb{R}^{T\times D}$ are passed through multi-head attentive temporal pooling, with keys derived from $\bar{\mathbf{K}}$ and queries and values from $\bar{\mathbf{V}}$. A lightweight Refiner module and embedding MLP then map the pooled representation to a speaker embedding $\mathbf{e}\in\mathbb{R}^{256}$.

The head is optimized with ArcFace loss \cite{deng2019arcface} (margin $m{=}0.3$, scale $s{=}30$),
\begin{equation}
\mathcal{L}_{\mathrm{ASV}}=-\log\frac{e^{\,s\cos(\theta_{y}+m)}}{e^{\,s\cos(\theta_{y}+m)}+\sum_{j\neq y}e^{\,s\cos\theta_{j}}},\quad
\cos\theta_{j}=\hat{\mathbf{w}}_j^{\top}\hat{\mathbf{e}},
\end{equation}
where $\hat{\mathbf{e}}$ and $\hat{\mathbf{w}}_j$ are the L2-normalized embedding and class prototypes. During inference, the final ASV score is the cosine similarity $s_{\mathrm{ASV}}=\langle \hat{\mathbf{e}}_{\mathrm{enr}},\hat{\mathbf{e}}_{\mathrm{tst}}\rangle$ between the enrollment and the test embeddings.

\begin{table*}[t]
\centering
\caption{Results of all systems. All values are reported as EER (\%) except for a-DCF. The table is organized according to the \textit{Setting} column. The \textbf{Base} section presents a comparison between our best configuration of the proposed USE model and the baseline systems. The \textbf{Ablation} section analyzes the contribution of each individual component. \textbf{Scalability} reports the results obtained when using two heads for the ASV task. All results up to this section are reported on the SpoofCeleb dataset. The \textbf{ASVspoof5} section presents the performance of the final version of the model on the ASVspoof5 dataset.}
\label{tab:results}
\setlength{\tabcolsep}{8.5pt}
\renewcommand{\arraystretch}{1.1}
\resizebox{\textwidth}{!}{
\begin{tabular}{l l ll c cccccccc}
\toprule
Dataset & Setting & \multicolumn{2}{c}{Sub-System} & Params
& \multicolumn{4}{c}{Dev}
& \multicolumn{4}{c}{Eval} \\
\cmidrule(lr){3-4}
\cmidrule(lr){5-5}
\cmidrule(lr){6-9}
\cmidrule(lr){10-13}
 &  & ASV & CM & \#Learnable
& SASV & CM & ASV & a-DCF
& SASV & CM & ASV & a-DCF \\
\midrule

\multirow{9}{*}{SpoofCeleb}

& \multirow{2}{*}{Baseline}
& E2E SASV \cite{wu2025wildspoof} & -- & 31M
& 11.29 & 47.64 & 11.83 & 0.256
& 10.74 & 49.62 & 11.64 & 0.230 \\

& 
& WavLM-ECAPA-TDNN \cite{far2026spoofing} & W2V-AASIST & 320M
& 3.38 & 0.44 & \textbf{3.36} & 0.043
& 3.93 & 0.16 & 3.86 & 0.047 \\

\cmidrule(lr){2-13}

& USE (\textit{ours}) & W2VBert-ASV & W2VBert-CM & 7.0M
& \textbf{3.27} & \textbf{0.26} & 4.05 & \textbf{0.040}
& \textbf{3.24} & \textbf{0.03} & \textbf{3.65} & \textbf{0.038} \\

\cmidrule(lr){2-13}

& \multirow{5}{*}{Ablation}

& W2VBert-ASV & \ \  w/o Cross L. Att. & 3.2M
& 3.48 & 0.36 & 4.07 & 0.047
& 3.26 & 0.05 & 3.79 & 0.039 \\

& 
& \ \  w/o Refiner & \ \  w/o Refiner & 7.0M
& 4.05 & 0.27 & 5.73 & 0.054
& 3.91 & 0.15 & 4.51 & 0.055 \\

&
& \ \  w/o Multilayer & \ \  w/o Multilayer & 3.0M
& 4.60 & 0.42 & 5.45 & 0.059
& 4.93 & 0.28 & 5.51 & 0.058 \\

&
& \ \  w/o WP & \ \  w/o WP & 7.0M
& 4.24 & 0.48 & 5.02 & 0.056
& 3.56 & 0.15 & 4.13 & 0.044 \\

&
& \ \  w/o All Prompts & \ \  w/o All Prompts & 6.2M
& 8.67 & 1.33 & 9.19 & 0.134
& 7.40 & 0.42 & 8.47 & 0.086 \\

\cmidrule(lr){2-13}

& Scalability
& \{SC, VC2\} ASV Heads & W2VBert-CM & 8.7M
& 2.94 & 0.26 & 3.51 & 0.037
& 2.73 & 0.03 & 3.15 & 0.035 \\

\midrule

ASVspoof5 & USE (\textit{ours})
& W2VBert-ASV & W2VBert-CM & 7.0M
& 3.31 & 0.24 & 3.98 & 0.043
& 5.23 & 3.60 & 4.27 & 0.092 \\

\bottomrule
\end{tabular}
}
\end{table*}

\subsection{CM Head: Cross-Layer Attention with DSU}

The CM head differs from the ASV head mainly in layer fusion. At each frame $t$, the layer outputs are stacked as $\mathbf{S}_t=[\mathbf{H}^{(1)}_t;\dots;\mathbf{H}^{(N)}_t]\in\mathbb{R}^{N\times D}$ and combined with 4-head cross-layer attention,
\begin{equation}
\mathbf{F}_t=\frac{1}{N}\sum_{n=1}^{N}\Big[\mathrm{softmax}\!\Big(\tfrac{(\mathbf{S}_t\mathbf{W}_Q)(\mathbf{S}_t\mathbf{W}_K)^{\top}}{\sqrt{d_h}}\Big)(\mathbf{S}_t\mathbf{W}_V)\Big]_n,
\end{equation}
yielding the fused frame sequence $\mathbf{F}\in\mathbb{R}^{T\times D}$. Spoofing cues often appear at different abstraction levels across attacks, so this adaptive recalibration is better suited to CM than the static layer weighting used in the ASV head.

During training, Domain Shifts with Uncertainty (DSU) \cite{li2022uncertainty} is applied to $\mathbf{F}$ with probability $0.3$ before temporal pooling to perturb feature statistics and improve robustness to unseen attacks and codec mismatch. The sequence then passes through 8-head attentive temporal pooling, the same Refiner bottleneck module as the ASV head, and an embedding MLP that produces a 256-dimensional representation for binary bona fide/spoof classification. The head is trained with weighted cross-entropy \cite{lin2017focal} using class weights $\mathbf{w}=[10,1]$ for the bona fide and spoof classes. The larger weight on bona fide counters the spoof-heavy class imbalance and matches the 10 attack types in the SpoofCeleb training set.

\subsection{SASV Score Integration}

A challenge in integrated SASV systems is that ASV and CM subsystems produce scores on different scales: ASV cosine similarities range in $[-1, 1]$, whereas CM classification produces bounded posteriors. We therefore apply a sigmoid $\sigma(\cdot)$ to calibrate the ASV cosine similarity $s_{\mathrm{ASV}}$ to $[0, 1]$, while the CM head directly outputs the bona fide posterior probability $p_{\mathrm{bona}}=[\mathrm{softmax}(\mathbf{z})]_0 \in [0, 1]$, where $\mathbf{z}\in\mathbb{R}^{2}$ are the CM logits over the bona fide and spoof classes. The final SASV score is the direct product of these calibrated values:
\begin{equation}
s_{\mathrm{SASV}}=\sigma(s_{\mathrm{ASV}})\cdot p_{\mathrm{bona}}.
\end{equation}
Each branch is trained through its own prompts and head, and the two scores are combined only at this stage. At inference, ASV and CM execute two prompt-conditioned forward passes while sharing the single 600M backbone in memory. The sigmoid calibration maps both outputs to $[0, 1]$, and their product acts as an AND-gate that is high only when accepted by both tasks. Unlike cohort normalization \cite{matvejka2017analysis}, this rule is fixed, parameter-free, and trial-independent.


\subsection{The SASV Datasets}
\label{subsec:dataset}

In this work we used two SASV benchmarks. The primary benchmark for this study is the SpoofCeleb dataset \cite{jung2025spoofceleb}. 
This dataset is a large-scale benchmark derived from VoxCeleb1, preserving its in-the-wild acoustic variability such as background noise, reverberation, and channel distortions. The dataset contains 1,251 speakers and approximately 150k bona fide utterances. Spoofed speech is generated using 23 contemporary speech synthesis systems, resulting in roughly 2.4 million synthetic utterances.



We also utilized the ASVspoof5 corpus under its Track 2 settings \cite{wang2024asvspoof}, which targets the SASV task. This dataset is derived from the crowdsourced Multilingual LibriSpeech corpus, features a diverse set of 1,922 unique speakers, and contains approximately 680,000 utterances. 
In this challenging dataset, the evaluation set includes 32 sophisticated spoofing attacks containing adversarial samples, all subjected to 7 distinct codec conditions such as Opus and Speex. 

We report the results using the Equal Error Rate (EER) and $\min$ $a$-DCF metrics under the official Wildspoof 2026 and ASVspoof 2024 challenges \cite{wu2025wildspoof, wang2024asvspoof} with cost parameters ($C_{\text{miss}}=1$, $C_{\text{fa}}=10$) and class priors ($\pi_{\text{tar}}=0.9405$, $\pi_{\text{non}}=0.0095$, $\pi_{\text{spf}}=0.05$). 



\section{Experimental Results}

This section reports the experimental evaluation of the proposed USE framework on the official partitions of the SpoofCeleb dataset. The optimal system configuration identified in these experiments is subsequently trained and evaluated on the ASVspoof5 benchmark. Before detailing the proposed architecture, we first establish performance benchmarks using two baseline systems.

\begin{figure}[t]
    \centering
    \includegraphics[angle=0, scale=0.43]{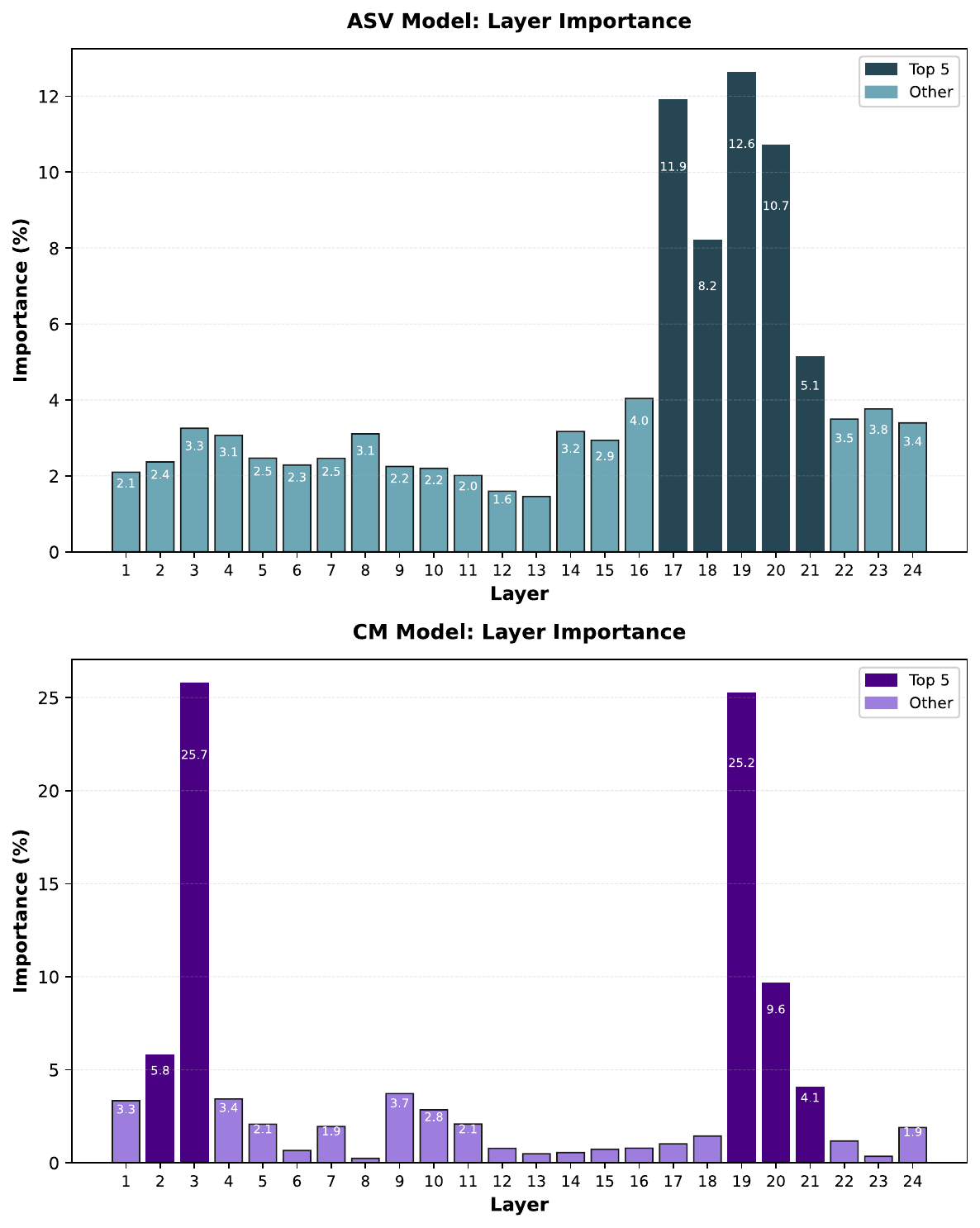}
    \caption{Learned layer weights of the 24 W2V-BERT layers for the ASV and CM heads. Darker colors indicate the five most important layers.}
    \label{fig:import_layer}
\end{figure}

\subsection{Baseline Systems}

As baseline systems, we implement two common SASV strategies consisting of an E2E architecture, which serves as the official WildSpoof challenge baseline on SpoofCeleb \cite{wu2025wildspoof}, and a cascade strategy comprising independent ASV and CM modules. 
 
In the E2E setup, we use SKA-TDNN \cite{mun2023frequency} as a single shared backbone to learn spoof-aware speaker embeddings by jointly training two output branches: a multi-class speaker identification branch for the ASV objective and a binary bona fide/spoof branch for anti-spoofing. This joint optimization tries to solve both tasks from the same representation, so the embedding space remains speaker-discriminative while being sensitive to spoof artifacts.
The system was trained with 400-frame chunks on the SpoofCeleb dataset. The model utilized margin-based softmax settings with a margin of 0.2 and a scale of 30. Optimization was performed using AdamW \cite{loshchilov2017decoupled} with a cosine annealing schedule and warmup over 50 epochs.
 As reported in Table \ref{tab:results}, this system achieves a min a-DCF of 0.256 on the development set and 0.230 on the evaluation set.


The second baseline consists of a competitive modular cascade system in which
the CM component, acting as the first module, utilizes a WPT-XLSR \cite{xuan2025wavesp} integrated with an AASIST \cite{jung2022aasist} back-end that receives features exclusively from the last layer of XLSR. This specific CM branch was trained using a weighted cross-entropy loss with a 1e-4 learning rate and a cosine annealing schedule for 15 epochs on the SpoofCeleb dataset. For the ASV component, we employed a pretrained WavLM-ECAPA-TDNN optimized on the VoxCeleb2 dataset \cite{chung2018voxceleb2}. 

As demonstrated in Table \ref{tab:results}, the performance of this system surpasses the baseline E2E system, achieving a min a-DCF of 0.043 on the development set and 0.047 on the evaluation set. However, training this configuration necessitates substantial computational resources to accommodate two SSL models, totaling 320 million trainable parameters. Furthermore, the use of distinct back-ends for each task requires specialized optimization for each component. A key limitation of this approach is its lack of modular scalability, where incorporating new data or additional task-specific heads is not feasible without the introduction of an entirely new backbone.



\subsection{USE Configuration}


By maintaining a frozen backbone in the proposed USE, we employ a parameter-efficient design for both tasks. The ASV branch incorporates 23 prompts across all 24 layers, with the task head totaling 1.5M trainable parameters.
The CM head follows a similar workflow but utilizes a cross-layer attention mechanism instead of a weighted sum, resulting in 5.5M trainable parameters. Both heads were trained independently for 10 epochs using AdamW optimization with a learning rate of 1e-4, a batch size of 16 on SpoofCeleb, and a fixed audio length of 64,600 samples, without joint multi-task backpropagation.

As illustrated in Figure \ref{fig:import_layer}, the learned layer weights reveal distinct strategies for the ASV and CM tasks. The ASV head demonstrates a complex reliance on the full depth of the backbone, and a significant weight assigned to layers 17 through 21. This suggests that identity verification primarily benefits from the high-level semantic features captured in the deeper stages of the model. In contrast, the spoof detection task utilizes a sparser distribution, focusing more on layers 2, 3, 19, 20, and 21. This indicates that optimal discrimination between spoofed and bona fide speech requires a multi-scale approach, combining low-level spectral patterns from early layers with long-range dependencies extracted from the final layers.

The proposed system, despite tuning only around 1\% of total parameters, achieves a state-of-the-art EER of 0.03\% for CM on this dataset. The ASV head achieves a 3.65\% EER, culminating in a final SASV performance of 0.038 min a-DCF and 3.24\% SASV-EER. When compared to the official SpoofCeleb leaderboard \footnote{\href{https://wildspoof.github.io/91_results/}{WildSpoof Challenge Results Page}}, our USE framework ranks as the second-best performing system.

\subsection{Ablation and Scalability}
\label{subsec:ablation}

To evaluate the contribution of each component, we present an ablation study in Table \ref{tab:results}. Replacing cross-layer attention with a weighted sum in the CM head reduces total trainable parameters from 7.0M to 3.2M with only marginal degradation (eval min a-DCF moving from 0.038 to 0.039). Removing the Refiner module causes eval min a-DCF to degrade to 0.055, and omitting multi-layer integration causes SASV-EER to drop to 4.93\% (Eval). Finally, learnable prompts provide the largest gain: removing all prompts causes eval min a-DCF to drop to 0.086; regular prompts alone recover it to 0.044, while adding wavelet prompts (USE) further improves it to 0.038 (and CM-EER from 0.15\% to 0.03\%) at the same 7.0M parameter budget.


\begin{table*}[t]
\centering
\caption{CM EER (\%) per attack type on the ASVspoof5 evaluation set.}
\label{tab:eval_attack_eer}
\resizebox{\textwidth}{!}{%
\begin{tabular}{l c c c c c c c c c c c c c c c c}
\toprule
System & A29 & A17 & A26 & A21 & A24 & A22 & A25 & A20 & A19 & A23 & A18 & A32 & A27 & A30 & A31 & A28 \\
\midrule
USE (ours) & 0.48 & 0.76 & 0.88 & 0.99 & 1.36 & 1.49 & 1.61 & 2.09 & 3.22 & 3.44 & 3.48 & 4.16 & 4.16 & 4.65 & 4.97 & 7.74 \\
\bottomrule
\end{tabular}%
}
\end{table*}

\begin{table}[t]
\centering
\caption{CM EER (\%) per codec on the ASVspoof5 evaluation set.}
\label{tab:codec_eer}
\begin{tabular}{l c c c c c c c}
\toprule
System & C02 & C10 & C01 & C03 & C09 & C08 & C11 \\
\midrule
USE (ours) & 2.88 & 3.05 & 3.13 & 3.19 & 3.32 & 3.49 & 4.48 \\
\bottomrule
\end{tabular}
\end{table}

Another critical advantage of the proposed USE architecture is its modular scalability. By leveraging WPT on a frozen backbone, the system allows for the independent addition of task-specific heads without altering the weights or functionality of existing modules. To evaluate this capability, we integrated a second ASV head with its 23 exclusive prompts, trained on the large-scale VoxCeleb2 dataset \cite{chung2018voxceleb2}, by adding only 1.5M learnable parameters. This auxiliary head followed the same hyperparameters as the original ASV module.

Following the training phase, the two ASV heads were integrated via score averaging and subsequently combined with the CM head using dot product. As demonstrated in Table \ref{tab:results}, this expanded configuration enhances system performance, yielding SASV-EERs of 2.94\% and 2.73\% for the development and evaluation sets, respectively. Furthermore, the final system achieved a 0.035 min a-DCF. We note that this configuration is presented as a demonstration of the framework's modularity rather than a like-for-like improvement over the base USE system: the auxiliary head is trained on out-of-domain VoxCeleb2 data, so the base and scaled settings are not directly comparable, and the gain should be read as evidence of scalable head expansion rather than of the core architecture alone.

\subsection{Evaluation on ASVspoof5 Dataset}
\label{sec:ASVspoof5}

ASVspoof5 is challenging due to its inclusion of diverse adversarial attacks and a wide array of transmission codecs. To ensure the framework's resilience, we incorporated all the 7 codecs of evaluation data as an augmentation strategy during the training phase with a per-utterance probability of $p=0.35$.


The performance of the proposed model is summarized in Table \ref{tab:results}. On the evaluation set, the system achieves a $\min$ $a$-DCF of $0.092$. When compared to the official ASVspoof5 challenge leaderboard \cite{wang2024asvspoof}, our USE framework ranks as the second-best performing system in the open condition and better than all systems in the closed condition. Remarkably, this performance was achieved without the use of additional training data beyond the provided corpus and by utilizing only a W2V-BERT 2.0 backbone. These results demonstrate the potential of our architecture to master multiple tasks on highly challenging, in-the-wild data through efficient parameter tuning and strategic data augmentation.

\subsection{ASVspoof5 Codec and Attack Breakdown}
\label{sec:asvspoof5_breakdown}

The aggregate CM-EER of 3.60\% on the ASVspoof5 evaluation set pools performance across seven transmission codecs and sixteen attack types. On SpoofCeleb, by contrast, the CM reaches 0.03\% EER on the evaluation set (Table~\ref{tab:results}), and we omit attack-wise CM breakdown there because spoof detection is already near ceiling on that benchmark and per-attack errors add little beyond the aggregate score. For ASVspoof5, Tables~\ref{tab:codec_eer} and~\ref{tab:eval_attack_eer} decompose the result to show where the countermeasure is reliable and where error remains concentrated.

Table~\ref{tab:codec_eer} reports CM-EER per evaluation codec. Errors stay between 2.88\% and 3.49\% for codecs C01--C10, with a larger increase to 4.48\% on C11, the most compressed condition. Codec variability therefore accounts for part of the eval CM error, but the spread across codecs is relatively narrow compared with the attack-wise breakdown.

Table~\ref{tab:eval_attack_eer} lists CM-EER for each evaluation attack. The model performs best on A29 and A17 (0.48\% and 0.76\%), while several generative attacks remain difficult, including A28 at 7.74\% and A31 at 4.97\%. The gap from below 1\% to nearly 8\% across attacks helps explain why the overall eval CM-EER is substantially higher than the SpoofCeleb result and suggests that hard attack subsets, rather than aggregate leaderboard placement alone, should guide further CM improvements.


\section{Conclusion}
\label{sec:conclusion}

In this paper, we presented a USE framework for robust SASV. By utilizing a W2V-BERT 2.0 backbone with WPT, we demonstrated that a parameter-efficient approach, tuning less than 1\% of the total parameters, can surpass the performance of resource-intensive cascaded and end-to-end architectures. Our analysis of the hierarchical feature wealth revealed that the ASV and CM tasks benefit from distinct layer-wise distributions, which our multi-layer aggregation strategy effectively captured.

The proposed system achieved a state-of-the-art CM-EER of 0.03\% on the SpoofCeleb dataset and a competitive min $a$-DCF of 0.035 in its expanded modular configuration. Furthermore, when evaluated on the challenging ASVspoof5 dataset, our framework achieved 0.092 as min a-DCF without requiring external training data. These results, combined with the architecture's modular scalability and small memory footprint, position the USE framework as an efficient and effective solution for deploying secure, multi-task biometric systems. The code and models are publicly available at \url{https://github.com/areffarhadi/unified-SASV}.



\section*{Acknowledgments}
Generative AI tools were used for writing assistance, including
language editing and revisions of the manuscript. All technical content and
conclusions were developed by the authors.

\bibliographystyle{IEEEtran}
\bibliography{mybib.bib}

\end{document}